\documentclass[twocolumn,pra,nofootinbib]{revtex4}
\usepackage{amsmath,amsfonts}
\usepackage{graphicx}

\def\be{\begin{equation}}
\def\ee{\end{equation}}
\def\bea{\begin{eqnarray}}
\def\eea{\end{eqnarray}}
\def\nn{\nonumber}

\begin{document}
\title{\bf Electric field on nucleus due to phonon lattice oscillations
in solid states}

\author{V.V. Flambaum and I.B. Samsonov}
\affiliation{$^1$School of Physics, University of New South Wales,
Sydney 2052,  Australia}

\begin{abstract}

In atoms and molecules, the electrons screen the nucleus from
the external electric field. However, if the frequency of the
electric field reaches the energy of atomic or molecular
transition, the electric field at the nucleus may be resonantly
enhanced by many orders in magnitude. In this paper, we study the
mechanisms of screening or enhancement of electric field acting on
the nuclei in solid states. We show that in dielectric crystals
the phonon oscillations of the lattice play crucial role for
determining the electric field on nuclei in the MHz-THz region. As
an application, we propose an experimental scheme for measuring
nuclear electric dipole moment in a solid state based on the
coherent excitation of acoustic phonons.

\end{abstract}

\maketitle

\section{Introduction}

According to Schiff's theorem \cite{Schiff}, atomic
electrons screen the nucleus from the external electric field.
In particular, a static electric field is
completely screened inside neutral atoms and molecules that
creates complications in the measurements of nuclear electric
dipole moments (EDMs). However, in real atoms and molecules this screening
is incomplete due to the magnetic moment effect \cite{Schiff,PGF}.
Moreover,
atomic EDM can be generated by the
nuclear Schiff moment and
magnetic quadrupole moment \cite{Schiffmoment1,Schiffmoment2,quadrupole}.

In ions and charged molecules the Schiff screening is incomplete
\cite{Dzuba,FlambaumKozlov}, but charged particles accelerate in the electric
field and quickly escape. Therefore, EDM measurement
experiments with ions require special configurations of the
electric and magnetic fields which trap the charged particles \cite{JILA}.

Another interesting possibility to circumvent the Schiff's screening
is to apply oscillating electric field. As is demonstrated in the
recent paper \cite{Flambaum2018}, the oscillating electric field
inside atoms is screened only partly, and the suppression factor is
proportional to the dynamic atomic polarizability. From the
physical point of view, we understand this result as a lag in the
displacement of the electron cloud in the changing electric field.
However, when the frequency of the electric field reaches one of
the energy levels of atomic transitions, the electric
field at nucleus may be enhanced by many orders in magnitude
\cite{Resonant}.

In molecules, the pattern of screening or resonant enhancement of the
oscillating electric field is much reacher
owing to the presence of rotational and vibrational
states in addition to the electronic states \cite{Mol}. Since the
energies of ro-vibrational states are lower than the
ones for electronic states, even the fields in the microwave
region may strongly interact with nuclei.

This paper is devoted to the extension of our previous results for
atoms \cite{Flambaum2018,Resonant} and molecules \cite{Mol} to the case of solid states. On the one
hand, a crystal can be viewed as a molecule composed of a
large number of atoms, and the general consideration developed in
Ref.\ \cite{Mol} is applicable. However, the general formulae in Ref.\ \cite{Mol} do not
allow us to give specific numerical estimates as one needs to know
exact wavefunctions which take into account positions of all atoms
in the solid state including all excitations. On the other hand,
in dielectrics and semiconductors, the electric field inside the
solid state is well understood through the electric permittivity
$\varepsilon = (1+\chi)\varepsilon_0$, where $\chi$ is the
electric susceptibility and $\varepsilon_0$ is the vacuum
permittivity. Indeed, the electric field ${\bf E}$ creates the
polarization in the medium ${\bf P} = \varepsilon_0\chi {\bf E}$
which reduces the electric field in the solid by the factor of
$\varepsilon_{\rm r} = 1+\chi$. However, this electric
field is macroscopic while we need to know the microscopic
electric field acting on each nucleus in the solid state.

In this paper, we consider solid states which
are represented by dielectric crystals with no free electrons
since the free electrons in metals are responsible for strong screening
which is unwanted in our case.
As we will demonstrate, it is
possible to estimate the electric field at each nucleus when the
phonon excitation of the lattice is known. For our estimates, it
will be sufficient to consider phonon lattice oscillations
semiclassically. Moreover, we will restrict ourselves to the case
of harmonic crystal assuming that anharmonic corrections are
subleading.

To sketch the main idea, consider a crystal in a state with no phonons such that the atoms
in the lattice do not vibrate. This means that each nucleus sits in
the minimum of the electrostatic potential of the lattice such
that the electric field at the points of the nuclei is vanishing.
Assume now that a phonon is excited with a given wavevector $\bf
q$ and frequency $\omega$. Then, for small perturbations, each atom (or ion) in the
lattice oscillates around its equilibrium position harmonically,  ${\bf u}(t) = {\bf u}_{0}
\cos \omega t$, where ${\bf u}$ is the deviation of the given
atom from its equilibrium position. Discarding the magnetic interaction, one can immediately
deduce that each nucleus with mass $m$ and charge $Ze$ accelerates in the electric field ${\bf
E}_{\rm nucl}= \frac{m}{Ze}\ddot {\bf u}$. Of course, at low
temperature, this electric field is very small as the amplitudes
of oscillations of atoms in the lattice are tiny. However, the
magnitude of this field becomes significant when the lattice
vibration is represented by a large number of phonons, e.g., near
the melting temperature of the crystal.\footnote{In liquids,
longitudinal sound waves also produce the electric field at
nuclei. However, in this paper we focus on solids only.}

At high temperature, the lattice contains large number of phonons of all allowed
frequencies and wavevectors. In this case, the oscillation of each
atom in the lattice is chaotic, with no preferred direction.
However, modern technology allows one to coherently create a given number of
phonons in the solid state with fixed polarization and frequency,
see, e.g., \cite{Gusev,sound}. Thus, when the pattern of the phonons in the solid state is
known, one can deduce the value of the oscillating electric field
acting on each nucleus. This technique may be useful in the
experiments aiming to study properties of atomic nuclei and, in
particular, the nuclear EDM.

In this paper, unless other units are explicitly specified, we use natural units, in which $\hbar=c=1$, where
$c$ is the speed of light.

The rest of the paper is organized as follows. In the next section
we derive analytical expressions for the electric
field at nucleus induced by acoustic and optical phonons in solids.
In section \ref{secnumer}, we estimate numerically the largest magnitude
of the electric field at nucleus in solid xenon created by
coherent acoustic phonons. We also consider the optical phonons in
a sodium chlorine crystal and show that they may enhance the
external electric field acting on the nucleus when the frequency
of this field is in resonance with the frequency of transverse
optical phonons. In section \ref{Appl}, we present a novel theoretical idea how
the coherently excited acoustic phonons may be used in
experiments aimed at measuring the permanent nuclear EDM. In the last
section we summarize and briefly discuss the obtained results.

\section{Theory}
\label{sectheory}
In this section, we derive analytical expressions for the value
of the electric field acting on nuclei in crystals due to acoustic
and optical phonons. In our derivation we restrict ourselves to
the theory of harmonic crystals leaving the analysis of anharmonic
corrections for further studies.

\subsection{Electric field due to acoustic phonons}


Acoustic phonons with long wavelengths may be considered as sound waves
in a continuous medium with a frequency $\omega$ and a wave vector ${\bf q}$,
\be
{\bf u}({\bf r},t) = {\bf u}_0 \cos({\bf q}\cdot{\bf r}-\omega t)\,,
\label{u}
\ee
where ${\bf u}_0$ is the amplitude. Consider, in particular, the
oscillations of the atom at ${\bf r}=0$, in the direction ${\bf u}_0=(0,0,u_0)$,
$u(t)=u_0 \cos\omega t$. This oscillation reaches the
maximum amplitude when the crystal is nearly melted or destroyed. Typically, this
oscillation amplitude is of order 0.1 of interatomic distance
$a$ \cite{SSP},
\be
u_{0,\rm max}\approx 0.1 a\,.
\ee
Thus, in the harmonic approximation, we know the
oscillation pattern of the atomic nucleus,
\be
u(t) = 0.1 a \cos\omega t\,.
\ee
This motion is produced by the
microscopic electric field acting on nucleus,
\be
E_{\rm nucl} = \frac{m}{Ze} \ddot u
=-\frac{0.1 a m\omega^2}{Ze}\cos\omega t\,,
\ee
where $m$ is the mass of the nucleus and $Ze$ is its charge. Thus,
the maximum amplitude of the electric field on nucleus due to
acoustic phonons reads
\be
E_{0} = -\frac{0.1 a m\omega^2}{Ze}\,.
\label{E0max}
\ee
As we will show in section \ref{secXe}, for low frequencies this
electric field may be much stronger than the screened external
electric field.

\subsection{Electric field due to optical phonons}

Let us consider a simple cubic lattice with two ions per unit
cell. Let $({\bf u}_1,m_1)$ and $({\bf u}_2,m_2)$ be pairs of
coordinates and masses of the two ions in the unit cell. To
describe the oscillations around the common center of mass, it is
convenient to introduce the phonon coordinate
\be
{\bf u} = \sqrt{m N} ({\bf u}_1-{\bf u}_2)\,,
\ee
where $m$ is the reduced mass, $m^{-1} = m_1^{-1}+m_2^{-1}$, and
$N$ is the atom number density. In the external oscillating electric field
${\bf E}={\bf E}_0 \cos\omega t$ the dynamics of the variable $\bf u$ is
described by the equation
\be
\ddot {\bf u} + \gamma \dot{\bf u} + \omega_{\rm t}^2 {\bf u} =
Z^* {\bf E}_0 \cos\omega t\,,
\label{w-eq}
\ee
where $\gamma$ is the phonon inverse lifetime, $\omega_{\rm t}$ is
the transverse optical phonon frequency and $Z^*$ is the
effective ion charge. The latter may be expressed via the
dielectric constants of the solid state (see, e.g., \cite{SSP}),
\be
Z^* = \omega_{\rm t} \sqrt{\frac{\varepsilon_{\rm stat} - \varepsilon_{\rm opt}}{4\pi}}
\frac{3}{\varepsilon_{\rm opt}+2}
\,,
\ee
where $\varepsilon_{\rm stat}$ and $\varepsilon_{\rm opt}$ are
the electric permittivity for static and high-frequency electric
fields, respectively.

In the harmonic approximation,
both $\omega_{\rm t}$ and $Z^*$ are constants.
More generally, one can consider the phonon dynamics with
anharmmonic corrections, but this problem is beyond the scope of
this paper.

The steady state solution of the equation (\ref{w-eq}) reads
\be
{\bf u} = {\bf u}_0 \cos(\omega t + \phi)\,,
\ee
where
\bea
{\bf u}_0&=& -\frac{Z^*{\bf E}_0}{
\sqrt{(\omega_{\rm t}^2-\omega^2)^2+(\omega\gamma)^2}}\,,\\
\phi &=&\arctan\frac{\omega\gamma}{\omega^2-\omega_{\rm t}^2}\,.
\eea
The electric field acting on the ions can be found as
\be
{\bf E}_{\rm ion} = \frac1q \sqrt{\frac m N} \ddot{\bf u}
=-\frac{\omega^2}{q}\sqrt{\frac m N} {\bf u}_0\cos(\omega
t+\phi)\,,
\ee
where $q$ is the charge of the ion. The amplitude of this field is
\bea
{\bf E}_{\rm ion,0} &=&-\frac{\omega^2}{q}\sqrt{\frac m N} {\bf
u}_0\label{Egeneral}\\
&=&\frac{Z^*}q\sqrt{\frac mN}
\frac{\omega^2 }{\sqrt{(\omega_{\rm t}^2-\omega^2)^2 + (\omega\gamma)^2}}
{\bf E}_0\,.\nn
\eea

When the frequency of the external electric field $\omega$ is far from the
phonon resonance frequency $\omega_{\rm t}$, $\omega\ll\omega_{\rm t}$, the equation
(\ref{Egeneral}) reduces to
\be
{\bf E}_{\rm ion,0} =\frac{Z^*}q\sqrt{\frac mN}
\frac{\omega^2 }{\omega_{\rm t}^2-\omega^2}
{\bf E}_0\,.
\label{E1}
\ee
On the contrary, near resonance, $\omega\approx\omega_{\rm t}$, $|\omega-\omega_{\rm t}|\ll
\gamma$, the amplitude of the electric field amplifies,
\be
{\bf E}_{\rm ion,0} =\frac{Z^*}q\sqrt{\frac mN}
\frac{\omega_{\rm t}}{\gamma}
{\bf E}_0\,.
\label{E2}
\ee

Finally, we point out that the bound electrons of each ion further
screen the electric field on the nucleus by the law
\cite{Flambaum2018}:
\be
{\bf E}_{\rm nucl} = {\bf E}_{\rm ion}
\left(
1-\frac{N_e}{Z} - \alpha_{\rm ion}(\omega) \frac{\omega^2 m_{e}^2}{Ze^2}
\right)\,,
\label{atom-screeniing}
\ee
where $N_e$ is the number of electrons in the ion, $Z$ is the
charge of the nucleus and $\alpha_{\rm ion}(\omega)$ is the atomic
polarizability of the ion.

\section{Numerical estimates}
\label{secnumer}
In this section, we give numerical estimates of the electric field
on nucleus induced by acoustic and optical phonons. For the
acoustic phonons, we consider solid xenon because it may be
further applied in nuclear EDM measurement experiments. The electric field
due to the optical
phonons is considered in the sodium chlorine crystal because its
optical and dielectric properties are well represented in the literature.

\subsection{Electric field in solid xenon}
\label{secXe}

Let us consider $^{129}$Xe noble gas in the solid state below
the melting temperature $T_{\rm melt} = 161$ K. The nuclear magnetic
dipole moment of this isotope is $\mu \approx -0.78 \mu_N $
\cite{muXe}, where $\mu_N$ is the nuclear magneton. In a strong
magnetic field $B=10$ T, the nuclear spin Larmor precession
frequency is $\omega= 2\mu B = 119 \mbox{ MHz } = 4.9\times 10^{-7}$
eV. We will be interested in acoustic phonons in the xenon
crystal with this frequency $\omega$.

The crucial assumption in our estimate will be that it is possible
to {\it coherently} excite the phonons in the solid state with the
given frequency and the wave vector. In particular, it is possible
to use a pulsed laser with pulses modulated to a given
frequency $\omega$ to coherently generate acoustic phonons,
see, e.g. \cite{Gusev}, for a review.
Moreover, we point out that the modern technology allows one to
measure precisely the number of phonons created in the solid state
\cite{sound}. Without going further into the details of these
techniques we will assume that they may be applied to a
solid xenon sample.

Given that the lattice parameter in the solid xenon is
$a=6.2\ \AA$, the maximum electric field at the nucleus may be
estimated from Eq.\ (\ref{E0max}):
\be
E_{0,\rm max} \approx 0.9  \mbox{ V/m }.
\label{Eest}
\ee

It is instructive to compare the electric field due to phonon
lattice vibrations with the screening of the external electric field
on the nucleus in an isolated Xe atom. According to
\cite{Flambaum2018}, the electric field at the nucleus induced by
the external electric field $E=E_0 \cos(\omega t)$ is
\be
E_{\rm nucl} = \frac{\tilde \alpha \tilde \omega^2}{Z} E
\approx 1.6\times 10^{-16} E\,,
\label{Escreened}
\ee
where $\tilde \alpha = 27.3$ is the static atomic polarizability
of xenon and $\tilde\omega = \omega a_B/e^2\approx 1.8\times 10^{-8}$ is the energy in atomic
units. Thus, for any reasonable laboratory electric field $E_0$ the
electric field on the nucleus due to phonons (\ref{Eest}) is much
stronger than the external electric field screened by the
atomic electrons (\ref{Escreened}).

\subsection{Electric field due to optical phonons in
sodium chlorine}

Dielectric and phonon properties of NaCl are well known, see, e.g., \cite{SSP}.
In particular, the electric permittivity for the static and optical
frequency electric fields is $\varepsilon_{\rm stat}=5.9$, $\varepsilon_{\rm
opt}=2.34$, and the transverse optical phonon resonant frequency
is $\omega_{\rm t}=0.02$ eV. The phonon width $\gamma$ may be
estimates as $\omega_{\rm t}/\gamma \approx 50$ \cite{width}. The
reduced mass for the NaCl molecule is $m=13.9 m_p$, and the atom number
density is $N\approx 2\times 10^{22}$ cm$^{-3}$. According to Eq.\
(\ref{E2}), the electric field acting on each ion is
\be
E_{\rm ion,0} \approx 39 E_0\,.
\ee

Given the electric field at ion, the field at nucleus may be
obtained with Eq.\
(\ref{atom-screeniing}). At the frequency $\omega=\omega_{\rm
t}$ the last term in Eq.\ (\ref{atom-screeniing}) is negligible. Then, for the Na ion $Z=11$ and $N_e =
10$, and we have
\be
E_{\rm nucl} \approx \frac1{11} E_{\rm ion}
\approx 3.5 E\,.
\label{linE}
\ee
Thus, the oscillating electric field in resonance with the
transverse optical phonons not only reaches the nucleus, but may also
be enhanced by the factor of 3.5.

Note that, in Eq.\ (\ref{linE}), the electric field at the nucleus grows
linearly with $E$ only for a weak external field when the harmonic
description of phonons applies. We stress that the magnitude of the electric field
cannot exceed the value (\ref{E0max}) at which the lattice
oscillations reach the maximum amplitude. For the NaCl crystal, this maximum
electric field is of order $10^7$ V/m at the frequency  $\omega=\omega_{\rm
t}=480$ GHz.

Note also that if one considers the electric field off the resonance
with the optical phonon frequency, the equation (\ref{E1})
applies. In this case, instead of the resonance enhancement there
is off-resonance suppression.

\section{Application: Nuclear EDM measurement}
\label{Appl}

Recently, there have been proposed a few experiments to measure
nuclear or electron EDMs with solid state samples
\cite{ex0,ex1,ex2,ex3,ex4}. It is expected that such experiments
may have a better sensitivity as compared to the traditional EDM
experiments on atoms and molecules in vapor state or in beams.
However, as was noticed in \cite{ex2,CASPER}, one of the
issues of the solid-state EDM experiments is that the
oscillating strong electric field causes the heating of the sample
and thermal depolarization of spins. However, this problem does
not arise in the CASPEr experiment \cite{CASPER,CASPER1} aimed at the
detection of oscillating EDM induced by axion dark matter.

\begin{figure}[h]
\begin{center}
\includegraphics[width=6cm]{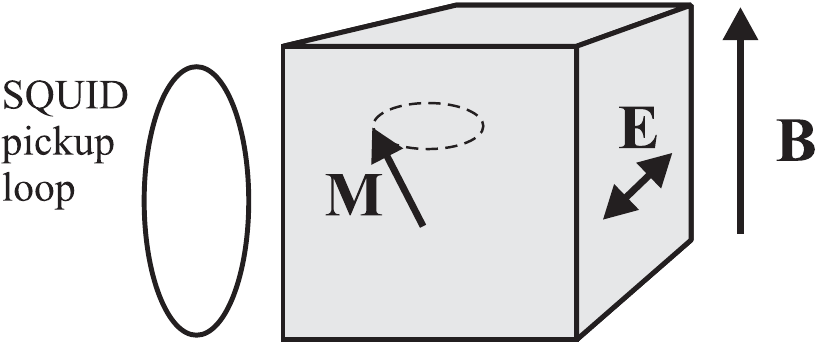}
\end{center}
\caption{
Orientation of magnetic and electric fields in the solid-state EDM experiments.
The external magnetic field $\bf B$ is created by a strong magnet while the electric field
$\bf E$ is the microscopic electric field on nuclei created by coherent phonon
lattice oscillations in the given direction. The interaction of nuclear EDM with
oscillating electric field creates the transverse macroscopic magnetization $\bf M$ of the sample
which may by detected by a SQUID magnetometer.}
\label{fig1}
\end{figure}

In this section, we will present a novel idea how to measure
permanent nuclear EDM using the phonon excitations in solids.
We will consider an experimental setup similar to
the CASPEr experiment \cite{CASPER,CASPER1}, but with important
modifications needed to measure permanent nuclear EDM. A very
schematic design of this experiment with main emphasis on the
orientation of electric and magnetic fields is shown in Fig.\ \ref{fig1}.
Nuclear spins in this solid state are pre-polarized by the
external strong magnetic field ${\bf B}$. The magnitude of this
field determines the Larmor precession frequency $\omega = 2 \mu
B$, where $\mu$ is the nuclear magnetic moment.

The oscillating electric
field $E_{\rm nucl}=E_{\rm nucl,0} \cos\omega t$ acting on the
nuclei should be created in the orthogonal plane. We assume
that this field is coherently created due to acoustic phonons as
described in section \ref{secXe}. The frequency of this field
should match the Larmor precession frequency of nuclear spins. In
this case, the NMR-like interaction of the oscillating electric field with
the nuclear EDM $d_N$ creates the macroscopic magnetization $M$ of the sample
in the direction orthogonal to both $\bf B$ and ${\bf E}_{\rm nucl}$.
For a small $(d_N\cdot t)$ this magnetization grows linearly with
time,
\be
M(t)\approx N \mu E_{\rm nucl} d_N t \sin(2\mu B t)\,,
\label{M}
\ee
where $N$ is the number density of nuclear spins,
$\mu$ is the nuclear magnetic moment and $B$ is the
external magnetic field. This magnetization may be detected
by a magnetometer like SQUID.

For the solid state, it is convenient to use solid xenon due to a
number of reasons: (i) It has odd stable isotopes $^{129}$Xe and
$^{131}$Xe with nuclear spins 1/2 and 3/2, respectively,
which are often used in the EDM measurements with NMR technology
\cite{Allmendinger,2019,Rosenberry2001}. Here, for simplicity, we consider
$^{129}$Xe only although other options are possible. (ii) The melting
temperature of xenon is high compared to other noble gases. (iii) In
solid state, xenon possesses cubic lattice which makes it
simple for theoretical studies. (iv) Lastly, since all electron shells are closed, only
nuclear EDM interacting with the electric field may be responsible
for the macroscopic magnetization of the sample.

For the solid xenon, the parameters in Eq.\ (\ref{M}) are: $N=1.66\times
10^{22}$ cm$^{-3}$, $\mu=-0.78\mu_N$, and $2\mu B=4.9\times
10^{-7}$ eV for $B=10$~T. The maximum electric field at nucleus
due to phonon lattice oscillations is estimated in Eq.\
(\ref{Eest}), $E_{\rm nucl}\approx0.9$ V/m.

The macroscopic magnetization of the sample may be detected with a
magnetometer. We assume a SQUID magnetometer with sensitivity
$10^{-15}$ T/$\sqrt{\rm Hz}$. Assuming that the measurement time may be of order
$t\sim 1000$ s (see, e.g., \cite{Romalis}), the solid state EDM experiment may detect the
permanent nuclear EDM of order
\be
d_N \approx 1.2 \times 10^{-22} \ e\cdot {\rm cm}.
\label{d}
\ee

The above estimate is very close to current experimental limits on the nuclear
EDM. The best limit on the atomic EDM in xenon was obtained in
Ref. \cite{Allmendinger}: $|d_{\rm Xe}|< 1.5\times 10^{-27}$
$e\cdot$cm. According to Ref.\ \cite{PGF}, in xenon, the nuclear
EDM contributes to the atomic EDM as $d_{\rm Xe} =4.4\times 10^{-6}
d_N$. Thus, the xenon EDM experiment puts the following limit on the
nuclear EDM in xenon:
\be
|d_N| < 2.5\times 10^{-22} \ e\cdot{\rm cm}.
\label{dlimit}
\ee
As a result, it is may me promising to develop an experimental
technique for measuring nuclear EDM using the phonon lattice
oscillations.

We stress that Eq.\ (\ref{d}) represents a very crude estimate of
sensitivity of an experiment aimed at measuring the nuclear EDM in solids
using lattice oscillations. In a
more accurate estimate one is to analyse the signal to
noise ratio by taking into account different noise sources in
realistic experiments. This analysis will be done elsewhere. Here
we only present the theoretical idea of possible application of
coherent phonon excitations in solids to the nuclear EDM
measurement. We point out that our proposal is novel as it allows for
{\it direct} nuclear EDM measurement which is free from the problem of
Schiff's screening.

Finally, we point out that the experimental technique described in
this section may be applied to ionic crystals. In this case, it
might be possible to measure the EDMs of ions induced by the
nuclear Schiff moment, magnetic quadrupole moment or electron EDM.

\section{Summary}

In this paper, we estimated the magnitude of the electric field
induced on the atomic nuclei by phonon lattice oscillations in
solid states. In these oscillations, the maximum deviation of
atoms from their equilibrium positions is typically of order of one
tenth of the interatomic distance that corresponds to the
maximum magnitude of the electric field acting on the nuclei as in
Eq.\ (\ref{E0max}). If the atoms oscillate randomly due to the thermal
motion, this filed averages out in time. However, if the acoustic
phonons can be created coherently in all atoms in the solid state
with a given wave vector and polarization, this electric field
becomes significant. As we demonstrated, for low frequencies
(characteristic to the acoustic phonons) this electric field acting on
atomic nuclei due to phonon lattice oscillation is much stronger
than the external electric field screened by the atomic electrons.
In particular, in solid xenon, this field is of order
1 V/m at the frequency 119 MHz.

The crucial assumption in our estimate is that the acoustic
phonons may be excited coherently with given wave
vector and polarization. This assumption is based on the advances in the
phonon generation and counting, see, e.g., \cite{Gusev,sound}. As
we advocate in section \ref{Appl}, this technique may be applied
to measure the nuclear EDM. We expect that possible NMR-like
experiments based on the phonon lattice oscillations may be
sensitive to the nuclear EDM of order $10^{-22}$ $e\cdot $cm. This
sensitivity is very close to the current constraint on the
nuclear EDM (\ref{dlimit}) arising from the recent atomic EDM measurement
in xenon \cite{Allmendinger}. Thus, it is tempting to
develop the experimental technique for measuring the nuclear EDM
with phonon lattice oscillations in solids.

We also estimated the electric field at nucleus due to
optical phonons which are present in many ionic crystals. In
contrast to the acoustic phonons, the optical ones interact
directly with the external electric field. When this electric
field is in resonance with the frequency of transverse
optical phonons, it may be enhanced on the nucleus.
Possible experimental applications of the electric field at
nucleus induced by the optical phonons will be discussed
elsewhere.

\subsection*{Acknowledgments}

This work is supported by the Australian Research Council
Grant No. DP150101405 and by a Gutenberg Fellowship.


\begin{thebibliography}{99}

\bibitem{Schiff} L.~I.~Schiff,
{\it Measurability of nuclear electric dipole moments},
Phys.\ Rev.\ {\bf 132}, 2194 (1963).

\bibitem{PGF} S.~G.~Porsev, J.~S.~M.~Ginges and V.~V.~Flambaum,
{\it Atomic electric dipole moment induced by the nuclear electric
dipole moment: The magnetic moment effect}, Phys. Rev. A {\bf 83},
042507 (2011).

\bibitem{Schiffmoment1} O.~P.~Sushkov, V.~V.~Flambaum,
  I.~B.~Khriplovich, {\it Possibility of investigating
  P- and T-odd nuclear forces in atomic and molecular
experiments},  Zh.\ Exp.\ Teor.\ Fiz.\ 87, 1521 (1984) [Sov.
  JETP 60, 873 (1984)].

\bibitem{Schiffmoment2} V.~V.~Flambaum, I.~B.~Khriplovich,
  O.~P.~Sushkov, {\it On the P-and T-nonconserving nuclear moments},
  Nucl.\ Phys.\ A {\bf 449}, 750 (1986).

\bibitem{quadrupole} V.~V.~Flambaum, {\it Spin hedgehog and collective magnetic
quadrupole moments induced by parity and time invariance violating
interaction}, Phys.\ Lett.\ B {\bf 320}, 211 (1994).

\bibitem{Dzuba}  V.~A.~Dzuba, V.~V.~Flambaum, P.~G.~Silvestrov, O.~P.~Sushkov,
{\it Shielding of an external electric field in atoms},
Phys.\ Lett.\ A {\bf  118}, 177 (1986).

\bibitem{FlambaumKozlov} V.~V.~Flambaum and A.~Kozlov,
{\it Extension of the Schiff theorem to ions and molecules}
Phys.\ Rev.\ A {\bf 85}, 022505 (2012).

\bibitem{JILA}
  W.~B.~Cairncross {\it et al.},
  {\it Precision measurement of the electron's electric dipole moment using trapped molecular ions},
  Phys.\ Rev.\ Lett.\  {\bf 119}, no. 15, 153001 (2017).

\bibitem{Flambaum2018}
  V.~V.~Flambaum,
  {\it Shielding of an external oscillating electric field inside atoms},
  Phys.\ Rev.\ A {\bf 98}, no. 4, 043408 (2018).

\bibitem{Resonant}
  V.~V.~Flambaum and I.~B.~Samsonov,
  {\it Resonant enhancement of an oscillating electric field in an atom},
  Phys.\ Rev.\ A {\bf 98}, no. 5, 053437 (2018).

\bibitem{Mol}
  H.~B.~T.~Tan, V.~V.~Flambaum and I.~B.~Samsonov,
  {\it Screening and enhancement of an oscillating electric field in molecules},
  Phys.\ Rev.\ A {\bf 99}, no. 1, 013430 (2019).

\bibitem{Gusev} P.~Ruello and V.~E.~Gusev, {\it Physical mechanisms of
coherent acoustic phonons generation by ultrafast laser action},
Ultrasonics {\bf 56}, 21 (2015).

\bibitem{sound} L.~R.~Sletten, B.~A.~Moores, J.~J.~Viennot, and
K.~W.~Lehnert, {\it Resolving Phonon Fock States in a Multimode Cavity with a Double-Slit
Qubit}, Phys.\ Rev.\ X {\bf 9}, 021056 (2019).

\bibitem{SSP} N.~W.~Ashcroft and N.~D.~Mermin, {\it Solid State
Physics}, Saunders College Publishing, 1976.

\bibitem{muXe} W. Makulski,
{\it $^{129}$Xe and $^{131}$Xe nuclear magnetic dipole moments from
gas phase NMR spectra}, Magn.\ Reson.\ Chem.\ {\bf 53}, 273 (2015).

\bibitem{width} G.~Raunio,
{\it Phonon widths in Nacl, KCl, and RbCl from neutron
measurements}, Phys.\ Stat.\ Sol.\ {\bf 35}, 299
(1969).

\bibitem{ex0} S.~K.~Lamoreaux,
{\it Solid-state systems for the electron electric dipole moment and other fundamental measurements},
Phys.\ Rev.\ A {\bf 66}, 022109 (2002).

\bibitem{ex1} T.~N.~Mukhamedjanov and O.~P.~Sushkov, {\it Suggested search for
$^{207}$Pb nuclear Schiff moment in PbTiO$_3$ ferroelectric},
Phys.\ Rev.\ A {\bf 72}, 034501 (2005).

\bibitem{ex2} D.~Budker, S.~K.~Lamoreaux, A.~O.~Sushkov, and
O.~P.~Sushkov, {\it Sensitivity of condensed-matter P- and
T-violation experiments}, Phys.\ Rev.\ A {\bf 73}, 022107 (2006).

\bibitem{ex3} A.~O.~Sushkov, S.~Eckel, and S.~K.~Lamoreaux,
{\it Prospects for an electron electric-dipole-moment search with
ferroelectric (Eu,Ba)TiO$_3$ ceramics},
Phys.\ Rev.\ A {\bf 81}, 022104 (2010).

\bibitem{ex4} K.~Z.~Rushchanskii {\it et al.}, {\it A multiferroic material
to search for the permanent electric dipole moment of the
electron}, Nat.\ Mater.\ {\bf 9}, 649 (2010).

\bibitem{CASPER} D.~Budker, P.~W.~Graham, M.~Ledbetter, S.~Rajendran, and
A.~O.~Sushkov, {\it Proposal for a Cosmic Axion Spin Precession
Experiment (CASPEr)}, Phys.\ Rev.\ X {\bf 4}, 021030 (2014).

\bibitem{CASPER1}
  D.~F.~Jackson Kimball {\it et al.},
  {\it Overview of the Cosmic Axion Spin Precession Experiment (CASPEr)},
  arXiv:1711.08999 [physics.ins-det].

\bibitem{Allmendinger}
  F.~Allmendinger {\it et al.},
  {\it Measurement of the Permanent Electric Dipole Moment of the $^{129}$Xe Atom},
  Phys.\ Rev.\ A {\bf 100}, no. 2, 022505 (2019).

\bibitem{2019} N.~Sachdeva {\it et al}.,
{\it A new measurement of the permanent electric dipole moment
of $^{129}$Xe using $^3$He comagnetometry and SQUID detection},
arXiv:1902.02864v1 (2019).

\bibitem{Rosenberry2001} M.~A.~Rosenberry and T.~E.~Chupp,
{\it Atomic electric dipole moment measurement using spin
exchange pumped masers of $^{129}$Xe and $^3$He},
Phys.\ Rev.\ Lett. {\bf 86}, 22 (2001).

\bibitem{Romalis} M.~V.~Romalis and M.~P.~Ledbetter,
{\it Transverse spin relaxation in liquid $^{129}$Xe
 in the presence of large dipolar fields}, Phys.\ Rev.\ Lett.\ 87,
 067601 (2001).


\end{thebibliography}
\end{document}